\documentclass[aps,prl,twocolumn,superscriptaddress,showpacs,nobalancelastpage]{revtex4}

\usepackage{epstopdf}
\usepackage{graphicx}
\usepackage{dcolumn}
\usepackage{bm}
\usepackage{amsmath}
\usepackage{hyperref}
\usepackage[dvips]{color}

\newcommand{\be}{\begin{equation}}
\newcommand{\ee}{\end{equation}}
\newcommand{\beq}{\begin{eqnarray}}
\newcommand{\eeq}{\end{eqnarray}}

\def\nue{\mathrel{{\nu_e}}}
\def\numu{\mathrel{{\nu_\mu}}}
\def\nutau{\mathrel{{\nu_\tau}}}
\def\nux{\mathrel{{\nu_x}}}

\def\barnue{\mathrel{{\bar \nu}_e}}
\def\barnumu{\mathrel{{\bar \nu}_\mu}}
\def\barnutau{\mathrel{{\bar \nu}_\tau}}
\def\barnux{\mathrel{{\bar \nu}_x}}

\def \lta {\mathrel{\vcenter{\hbox{$<$}\nointerlineskip\hbox{$\sim$}}}}
\def \gta {\mathrel{\vcenter{\hbox{$>$}\nointerlineskip\hbox{$\sim$}}}}

\def\t13{\mathrel{{\theta_{13}}}}
\def\y12{\mathrel{{\tan^2 \theta_{12}}}}
\def\c2{\mathrel{{\chi^2 }}}

\newcommand{\df}{DSN$\nu$F}

\begin{document}


\title{The diffuse neutrino flux from supernovae: upper limit on the electron neutrino component from the non-observation of antineutrinos at SuperKamiokande}

\author{Cecilia Lunardini}
\affiliation{Institute for Nuclear Theory and University of Washington, Seattle, WA 98195 }%

 
\begin{abstract}
I derive an upper bound on the electron neutrino component of the
diffuse supernova neutrino flux from the constraint on the
antineutrino component at SuperKamiokande.  The connection between
antineutrino and neutrino channels is due to the similarity of the
muon and tau neutrino and antineutrino fluxes produced in a supernova,
and to the conversion of these species into electron neutrinos and
antineutrinos inside the star.  The limit on the electron neutrino
flux is $5.5 ~{\rm cm^{-2} s^{-1}}$ above 19.3 MeV of neutrino energy,
and is stronger than the direct limit from Mont Blanc by three orders of
magnitude.  It represents the minimal sensitivity required at future
direct searches, and is intriguingly close to the reach of the Sudbury Neutrino Observatory (SNO) and
of the ICARUS experiment.  The electron neutrino flux will have a lower
bound if the electron antineutrino flux is measured.  Indicatively,
the first can be smaller than the second at most by a factor of 2-3
depending on the details of the neutrino spectra at production.

\end{abstract}                            
 
\maketitle
  
The observation of the diffuse flux of neutrinos from core collapse
supernovae has become a concrete possibility in the last years.  If
achieved, it would represent a milestone of neutrino astronomy, as it
would allow faster progress with respect to searches for an individual
supernova signal, and would provide several tests of neutrino
properties, of the physics of core collapse and of the cosmological
rate of supernovae.

Currently, the best experimental result on  
the diffuse supernova neutrino flux (\df\ from here
on) is the  upper limit on its  electron antineutrino ($\barnue$) component, put by SuperKamiokande (SK) above 19.3 MeV of neutrino energy \cite{Malek:2002ns}.  This limit comes from
the non-observation of events due to $\barnue$ absorption on protons in water
above the background. 
At
90\% confidence level, it reads: 
\be \Phi_{\bar e}(E>19.3~{\rm MeV})<1.2 ~{\rm
cm^{-2} s^{-1}}~.
\label{sklim}
\ee This bound approaches the range of theoretical predictions of the
flux
\cite{Totani:1995dw}--\cite{Lunardini:2005jf},
suggesting that a positive signal may be seen in the near
future. Such signal will add to the information given by the $\barnue$ data from SN1987A \cite{Hirata:1987hu,Bionta:1987qt}. 

The effort to detect or constrain the electron neutrino ($\nue$)
component of the \df\ is at least as important as  the search in the
$\barnue$ channel. 
Indeed,
it has the potential to give the very first data on \emph{neutrinos}
(instead than antineutrinos) from supernovae, with crucial first tests of theory -- e.g., of neutrino transport in dense media -- in the neutrino channel.
In spite of its importance, however, the search of the $\nue$ component of the \df\ has
had less progress than that of $\barnue$. This is
because in water an exclusive (one flavor only) $\nue$ signal is not possible, and non-water experiments  are
forced to smaller volumes than SK, resulting in a lower
sensitivity. 
The best experimental information on the $\nue$ flux  is
the bound: \be \Phi_{ e}(25~{\rm
MeV}<E<50~{\rm
MeV})<6.8\cdot 10^{3}~{\rm cm^{-2} s^{-1}}~,
\label{mb}
\ee at 90\% confidence level, found in 1992 at the Liquid Scintillator Detector (LSD) of Mont Blanc  from the non-observation
of $\nue$ absorption on $^{12}{\rm C}$ \cite{Aglietta:1992yk}.
New, stronger  bounds will come from the Sudbury Neutrino Observatory (SNO)  \cite{Beacom:2005it} and from the upcoming experiment for Imaging Cosmic And Rare Underground Signals (ICARUS) \cite{Cocco:2004ac}.

In this paper I show that  the
$\nue$ diffuse flux is \emph{already} strongly restricted by the SK
result itself. Indeed, the limit on $\barnue$, Eq. (\ref{sklim}),
can be easily translated into an analogous result for $\nue$.
This new bound improves greatly on the LSD limit.  Clearly, it has the same information
content as the direct bound (\ref{sklim}), and so it is important
mainly because it defines the minimum sensitivity required for future
direct searches in the $\nue$ channel.

Briefly, what allows to place a limit on the $\nue$ flux from the result on
$\barnue$ is the combination of two elements. The first is the fact that the
fluxes of muon and tau neutrinos and antineutrinos ($\numu,\nutau,\barnumu,\barnutau$) produced inside the
star  are equal to first approximation.
The second  element is that neutrinos undergo flavor conversion
due to masses and mixings on their way from production to detection.
This implies that at least part of the detected $\nue$ ($\barnue$) flux
originates from the produced $\numu$ and $\nutau$ ($\barnumu$,
$\barnutau$).  Their origin from species that have similar fluxes at
production provides the connection between the
$\nue$ and $\barnue$ channels.
This same connection will also give a \emph{ lower} bound on the $\nue$ component of the \df\ if the $\barnue$ flux is detected, thus giving important guidance to direct searches of the $\nue$ flux.

 Some generalities are in order. 
 In a core collapse supernova neutrinos are present as a thermal gas,
right outside the core, from where they diffuse out carrying
away about 99\% of the ${\mathcal O} (10^{53})$ ergs of gravitational
energy that is liberated in the collapse. This energy is shared  among neutrinos and antineutrinos of the three flavors, $e,\mu,\tau$, in comparable amounts.

 Before flavor conversion, the energy spectrum of each flavor of neutrinos is expected to be
nearly thermal  with average energy of 10-20 MeV  \cite{Keil:2002in}. 
The non-electron species, $\numu$, $\nutau$, $\barnumu$ and
$\barnutau$, interact with matter only via neutral current exchange,
which is flavor independent and, at the lowest order in momentum transfer, parity
preserving \cite{Fukugita:2003en} .
This implies that $\numu$ and $\nutau$ ($\barnumu$ and
$\barnutau$) have equal fluxes and so they can be considered as a
single species, $\nux$ ($\barnux$).  The spectra of $\nux$ and
$\barnux$ are similar, with  differences  due to weak magnetism corrections
to the neutrino-nucleus cross section. These cause 
the $\nux$ to have a $\sim 7\%$ lower
average energy \cite{Horowitz:2001xf}.  One also expects softer spectrum of $\nue$ with
respect to $\barnue$, because the interaction of $\nue$ ($\barnue$)
with matter is dominated by capture on neutrons (protons) and matter
is highly neutronized deep inside the star.  In summary, the hierarchy
of average energies $E_{0 \bar x} \sim E_{0 x} \gta E_{0 \bar e} \gta E_{0 e}$
is expected at the point of decoupling of neutrinos from matter, and is supported by numerical calculations (e.g. \cite{Totani:1997vj,Burrows:2000tj,Liebendoerfer:2003es}).
Indicative values of the average energies are: $E_{0 e}
\sim 9 - 16 $ MeV, $E_{0 \bar e} \sim 12 - 18$ MeV, $E_{0 x}\sim 15 -
22$ MeV.

Between the production and the detection points, the neutrino fluxes are modified by two effects: the redshift of energy and the flavor conversion (oscillations) in the star and in the Earth. The latter is due
to the interplay of neutrino masses, flavor mixing and coherent
interaction of neutrinos with the medium (refraction).  
In cases where the conversion has no zenith nor energy dependence, the  diffuse fluxes of $\nue$ and of $\barnue$ in a detector in a given energy bin $E_1 - E_2$ have the form:
\beq
\Phi_e= p \phi_e + (1-p) \phi_x~,
\label{fluxes_simpl_e} \hskip 0.5truecm
\Phi_{\bar e}= \bar p \phi_{\bar e} + (1-\bar p) \phi_{\bar x}~,
\label{fluxes_simpl}
\eeq 
where $p$ ($\bar p$) is the  $\nue$ ($\barnue$) survival probability,  i.e.  the probability that a neutrino produced as $\nue$ ($\barnue$) 
is detected as $\nue$ ($\barnue$). Here $\phi_w$ is the diffuse flux  of $\nu_w$ ($w=e,\bar e,x,\bar x$) in the detector in absence of conversion.  It depends on  the redshift $z$, on the $\nu_w$ flux originally produced in a individual star, $d
N_{w} (E^\prime)/d E^\prime $, and on the comoving rate of supernovae, $R_{SN}$:
\be \phi_w\equiv \frac{c}{H_0}\int_{E_1}^{E_2} {d}E \int_0^{z_{ max}} \frac{{d}
N_{w} (E^\prime)}{{d} E^\prime} \frac{R_{ SN}(z) {d}
z}{\sqrt{\Omega_{ m}(1+z)^3+\Omega_\Lambda}}~
\label{short}
\ee (see e.g. \cite{Ando:2004hc}). 
  Here $E$ and
$E^\prime=(1+z) E$ are the neutrino energy at Earth and at the
production point; $\Omega_{ m}\simeq 0.3 $ and $\Omega_\Lambda\simeq
0.7$ are the fraction of the cosmic energy density in matter and dark
energy respectively; $c$ is the speed of light and $H_0$ is the Hubble
constant.

Generally, the conversion of neutrinos  depends on
$E$ and $E^\prime$, on the matter density profile along the neutrino
trajectory -- and thus on the arrival direction, for neutrinos crossing the
Earth -- and on the neutrino masses and mixings.  In particular, $p$ and
$\bar p$ are functions of the mixing
angles $\theta_{12}$ and  $\theta_{13}$
(assuming the standard parameterization of the mixing matrix,  e.g.
\cite{Krastev:1988yu}) and of the  sign of the
atmospheric mass squared splitting, $\Delta m^2_{31}$.    The $\theta_{12}$ angle is measured  to be 
\be
\tan^2 \theta_{12}\simeq 0.3 - 0.73
\label{th12}
\ee at 99.73\% C.L. \cite{Aharmim:2005gt}, while  $\theta_{13}$ is still unknown, with the bound $\sin^2
\theta_{13}\lta 0.02$ \cite{Apollonio:1999ae}.   For brevity,  I will refer to the intervals
$\sin^2\theta_{13}\lta 2 \cdot 10^{-6}$ and $\sin^2\theta_{13}\gta 3 \cdot 10^{-4}$ as
``small" and ``large" $\theta_{13}$ respectively. The sign of $\Delta m^2_{31}$ is also unknown; the two possibilities are called ``normal hierarchy" ($\Delta m^2_{31}>0$) and ``inverted hierarchy" ($\Delta m^2_{31}<0$) of the neutrino mass spectrum.

Three  scenarios of conversion in the star are most relevant here (see
Table \ref{tab}) \cite{Dighe:1999bi,Lunardini:2003eh,Lunardini:2001pb}:\\
(i) the \emph{ ``adiabatic normal"}, characterized by the normal  hierarchy
and large $\theta_{13}$.  In this scenario
neutrinos undergo resonant conversion \cite{Mikheev:1986if} at matter density of
$\sim 10^3~{\rm g \cdot cm^{-3}}$. Their propagation is adiabatic,
i.e.  without transitions between the eigenstates of the
Hamiltonian, resulting in $p=0$.  The propagation of antineutrinos
is adiabatic too, with $\bar p=\cos^2\theta_{12}$.  \\
(ii) the \emph{``adiabatic inverted"}, with the inverted  hierarchy   and large $\theta_{13}$.  Both neutrinos and antineutrinos propagate adiabatically, but the resonant conversion affects antineutrinos, so that one has $ p=\sin^2\theta_{12}$ and $\bar p=0$.\\
(iii) the \emph{``non-adiabatic"}, characterized by small $\theta_{13}$.
Here resonant conversion is made ineffective by the complete breaking
of adiabaticity, and one has $
p=\sin^2\theta_{12}$ and $\bar p=\cos^2\theta_{12}$ for any hierarchy.  \\
In these three cases, $p$ and $\bar p$ have very little dependence on
the neutrino energy, thus justifying the use of the expressions
(\ref{fluxes_simpl}).  For intermediate values of $\theta_{13}$ and
normal (inverted) mass hierarchy, things are as in case (iii)  with the difference that $p$ ($\bar p$)  
is energy-dependent and $0<p<\sin^2 \theta_{12}$ ($0<\bar p<\cos^2 \theta_{12}$).  
The survival probabilities receive small corrections from oscillations in the Earth \cite{Dighe:1999bi,Lunardini:2001pb,Takahashi:2001dc} and shockwave effects \cite{Schirato:2002tg}.  The former will be discussed briefly later; the latter are neglected as they are smaller than  1-2\% in the energy interval where the \df\ is largest, $E \sim 20 - 30   $ MeV \cite{Ando:2002zj}. I also neglect terms of order $\theta^2_{13}$ and higher in  $p$ and $\bar p$.

Now I derive the upper bound on the $\nue$ flux, $\Phi_e$. I consider high energy thresholds, $E_1 \gta 20$ MeV,  motivated by the cuts necessary for background reduction at existing $\nue$ and $\barnue$ detectors \cite{Malek:2002ns,Beacom:2005it}.
First, I find an expression that is always greater or equal than the ratio $\Phi_e/\Phi_{\bar e}$:
\be
\frac{\Phi_e}{\Phi_{\bar e}} \leq \frac{p \phi_e + (1-p) \phi_x}{(1-\bar p) \phi_{\bar x}}~.
\label{magg_1}
\ee This follows from Eq. (\ref{fluxes_simpl}) by dropping the
$\phi_{\bar e}$ term there.  Next, I use the fact that
$\phi_x/\phi_{\bar x}\leq 1$, because the
$\nux$ spectrum is slightly softer than the $\barnux$ one, and has
average energy at or below the threshold $E_1$ (see e.g. fig. 4 in ref. \cite{Horowitz:2001xf}). I also consider that
the ratio (\ref{magg_1}) is largest when $\bar p$ has its maximum
value $\bar p_{max}$.  Thus: \be \frac{\Phi_e}{\Phi_{\bar e} }\leq
\frac{1}{1-\bar p_{max}} \left[ 1 +p\left(\frac{\phi_e}{\phi_{\bar
        x}}-1 \right)\right]~.
\label{magg_2}
\ee It appears that, in general, an upper limit on
$\Phi_e/\Phi_{\bar e}$ depends on a variety of unknowns through the
factor $\phi_e/\phi_{\bar x}$.  However, the bound simplifies to \be
\frac{\Phi_e}{\Phi_{\bar e} }\leq \frac{1}{1-\bar p_{max}}~,~~~{\rm
  if}~p=0~{\rm or}~\phi_e/\phi_{\bar x}\leq 1~.
\label{simpleupper}
\ee 
%
From Table \ref{tab}  I have $\bar
p_{max}=\cos^2\theta_{12}+{\cal O}(10^{-2})$, with the second term
representing the (positive) correction due oscillations in the Earth. 
From Eqs. (\ref{sklim}), (\ref{th12}) and (\ref{simpleupper}) I get:
\beq
&&\frac{\Phi_e}{\Phi_{\bar e}} \lta  \frac{1}{\sin^2 \theta_{12}}+{\cal O}(10^{-2})\simeq 4.6~, 
\label{bestrat} \\
&&\Phi_{e}(E>19.3~{\rm MeV})\lta 5.5~{\rm cm^{-2} s^{-1}}~,
\label{bestlim}
\eeq
if $p=0$ or $\phi_e/\phi_{\bar x}\leq 1$.
Explicitly, Eqs. (\ref{simpleupper})-(\ref{bestlim}) hold 
under one or the other of the following conditions:\\
(a)  the adiabatic normal scenario is realized (large $\theta_{13}$, $p=0$).  \emph{No other assumptions} -- on the neutrino fluxes or on the supernova rate, etc. -- are necessary, and thus in this case the constraint (\ref{bestlim}) is robust.\\
(b)  any other oscillations scenario is realized, or the conversion pattern remains unknown, and we know that $\phi_e/\phi_{\bar x}\leq 1$.  The latter condition will be called ``$\nux$ dominance". 
It is generally valid at $E\gta 20$ MeV for neutrino fluxes motivated by theory;
I will return to this point in a moment.
\begin{table}
\begin{tabular}{|l|l|l|l|l|}
\hline
\hline
 character &  hierarchy & $p$ & $\bar p$  & $\Phi_e/\Phi_{\bar e}$\\ \hline \hline
adiabatic & normal  & 0  & $\cos^2\theta_{12}$ & 0.67 - 4.6 \\ 
 &   &   &  & ($<4.$6) \\ \hline
adiabatic & inverted  & $\sin^2 \theta_{12}$ & 0 & 0.39 - 1   \\  
&   &   &  & ($>0.39$ ) \\ \hline
fully  non-adiabatic & any  & $\sin^2 \theta_{12}$  &  $\cos^2\theta_{12}$ & 0.39 - 4.6 \\
&   &   &  & ( any ) \\ 
\hline
\hline
\end{tabular}
\caption{The survival probabilities $p$ and $\bar p$, and the ratio $\Phi_e/\Phi_{\bar e}$  for different combinations of mass hierarchy and degree of adiabaticity of the neutrino propagation, under the conditions   $\phi_e/\phi_{\bar x}\leq1$ and $\phi_{\bar e}/\phi_x\leq1$.  The numbers in brackets give the results  without those conditions. I used ${\rm
Max}(\phi_{\bar x}/\phi_{ x})= 1.5$. The small effects of the Earth matter and of shockwaves, as well as ${\cal O}(\theta^2_{13})$ terms are neglected.}
\label{tab}
\end{table}

Eq. (\ref{bestlim}) was obtained using
the smallest $\theta_{12}$ in Eq. (\ref{th12}) and includes the correction
due to Earth effects. This was estimated numerically following refs.
\cite{Lunardini:2001pb,Lunardini:2005jf} for a large variety of
neutrino spectra and supernova rate functions; it contributes to the
final result (\ref{bestlim}) by at most 5\%. About the
significance of the bound (\ref{bestlim}), using the likelihood
information from refs. \cite{Malek:2002ns,Aharmim:2005gt}  I find that
the inequality $\Phi_{\bar e}/\sin^2 \theta_{12}<5.5~{\rm cm^{-2} s^{-1}}$ holds at $98\%$
confidence level.
\begin{figure}[htbp]
  \centering
 \includegraphics[width=0.49\textwidth]{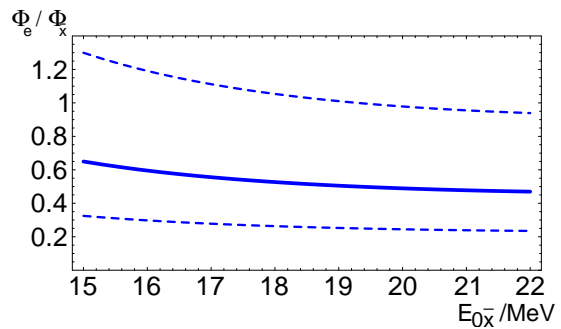}
\caption{The ratio $\phi_e/\phi_{\bar x}$ as a function of $E_{0\bar x}$, for $E_{0e}=12$ MeV and ratio of $\nue$ and $\barnux$ luminosities of 2,1,0.5 (uppper to lower curves).  I parameterized the neutrino spectra and the supernova rate function as in  \cite{Lunardini:2005jf}, with $\alpha_e=\alpha_{\bar x}=2.3$, $\alpha=0$, $\beta=3.4$. The SuperKamiokande energy threshold $E_1=19.3$ MeV was used.}
\label{ratioflux}
\end{figure}

A stronger bound follows from Eq. (\ref{simpleupper}) for the adiabatic inverted case ($\bar p_{max}=0$), with $\nux$ dominance:
\beq
\frac{\Phi_e}{\Phi_{\bar e}} \leq 1~, \hskip 0.4truecm
 \Phi_{e}(E>19.3~{\rm MeV})\lta 1.2~{\rm cm^{-2} s^{-1}}~.
\label{unitlim}
\eeq

The more general limit, Eq. (\ref{bestlim}),  is the main result of this work. 
It is highly motivating for $\nue$ detectors, as it approaches their sensitivities.  These were estimated to be $\sim 6 ~{\rm cm^{-2} s^{-1}}$  in the energy interval $22.5 -
32.5$ MeV  for  SNO (using charged
current $\nue$ scattering on deuterium) \cite{Beacom:2005it}, and $\sim 1.6 ~{\rm cm^{-2} s^{-1}}$ in the
interval $14 - 40$ MeV for ICARUS (using absorption in liquid Argon) \cite{Cocco:2004ac}.

One may wonder how Eq.  (\ref{bestlim}) changes when one
includes the dependence of neutrino conversion  on the
 energy, on the redshift and on the arrival direction of the
neutrinos. The derivation, and thus the 
results (\ref{magg_2}) and (\ref{simpleupper}), still apply, provided that for $p$, $\bar p$ and $\bar p_{max}$
one uses their values averaged  over energy,
redshift and zenith angle,  with the product of neutrino spectrum
and supernova rate as weight, see Eq. (\ref{short}).
This procedure gives the correction due to Earth
effects, explaining the ${\cal O}(10^{-2})$ term in Eq.
(\ref{bestrat}).  The limits (\ref{bestrat}) and (\ref{unitlim})
remain valid, as  they refer to limiting cases, where  $\bar p$ is
truly a constant.

A second question is the validity of the condition of $\nux$ dominance
above the threshold $E_1 \sim 20$ MeV. While there are no direct tests
of this, the condition is supported by current supernova simulations,
which predict the $\nue$'s to have comparable luminosity (within a
factor of two) and a softer spectrum than $\barnue$ and $\nux$.
Fig. \ref{ratioflux} gives the ratio $\phi_e/\phi_{\bar x}$ calculated
using Eq. (\ref{short}) with typical parameters (see caption). One
sees that the $\nux$ dominance is violated by at most $\sim 20\%$ when
the $\nux$ component is very soft and suppressed in luminosity.


What can be said on the $\nue$ flux  if evidence of the \df\ is not seen in the $\nue$ channel but is found in the
$\barnue$ one?  Similarly to what shown so far, a positive
detection of $\barnue$ can be translated into a {\it lower limit} on
the $\nue$ component.  Indeed, one can constrain the ratio $\Phi_{\bar
  e}/\Phi_e$ from above, using nearly the same steps as in
Eqs.(\ref{magg_1})-(\ref{unitlim}), with the exchanges $e
\leftrightarrow \bar e$, $x \leftrightarrow \bar x$, $p
\leftrightarrow \bar p$. The analogous of Eqs. (\ref{bestrat}) and (\ref{bestlim}) are:
\beq
\frac{\Phi_{\bar e}}{\Phi_{ e}} &&\lta  \frac{{\rm Max}(\phi_{\bar x}/\phi_{
x})}{\cos^2 \theta_{12}}+{\cal O}(10^{-2})~, \label{bestratupper} \\
\Phi_{ e}&&\gta   \Phi_{\bar e} \left[ 1.73~{\rm Max}\left( \frac{\phi_{\bar x}}{\phi_{ x}} \right)+{\cal
O}(10^{-2}) \right]^{-1}~,
\label{bestnuenum} 
\eeq 
where the largest value of $\theta_{12}$ from Eq. (\ref{th12}) was used. These results are
valid for the adiabatic inverted case ($\bar p=0$) or for $\nux$
dominance, $\phi_{\bar e}/\phi_{x}\leq1$.  They depend on the
maximum value allowed by theory of the ratio $\phi_{\bar x}/\phi_x$, ${\rm Max}(\phi_{\bar x}/\phi_{x})$,
just like the derivation of Eqs. (\ref{bestrat}) and (\ref{bestlim})
contains the maximum of $\phi_{x}/\phi_{\bar x}$. The latter was set
to 1 there.

To estimate ${\rm Max}(\phi_{\bar x}/\phi_{x})$  requires a dedicated work, with the evaluation of the dependence of $\phi_{\bar x}/\phi_x$ on the details of the $\nux$ and $\barnux$ original
spectra.
For illustration, here I take ${\rm
Max}(\phi_{\bar x}/\phi_{ x})= 1.5$, which is the largest ratio of the $\barnux$ and $\nux$ spectra found in ref. \cite{Horowitz:2001xf} in the energy interval  $E=20 - 80$ MeV.  With this value Eq. (\ref{bestnuenum}) gives $\Phi_{e} \gta
0.39 \Phi_{\bar e} -{\cal
O}(10^{-2})$ (the minus sign indicates a negative Earth effect).

For  the adiabatic normal scenario  ($p=0$) with the $\nux$ dominance one finds the analogous of Eq. (\ref{unitlim}):
\be
\frac{\Phi_{\bar e}}{\Phi_e} \leq {\rm
Max}(\phi_{\bar x}/\phi_{ x})~.
\label{unitlim2}
\ee
 This gives $\Phi_e\gta 0.67 \Phi_{\bar e}$ for  ${\rm
Max}(\phi_{\bar x}/\phi_{ x})=1.5$.  

The results of this paper are summarized in Table \ref{tab}.  The Table helps the discussion that follows. 

What will we learn from a detection of the \df\ in the $\nue$ channel, together with the limits (\ref{bestlim}) and (\ref{unitlim})?
If evidence is found above the value in (\ref{bestlim}), the physics
and/or the data analyses involved would have to be revised, e.g. by relaxing possible priors or by reconsidering background subtraction.
If both  the $\nue$  and the $\barnue$ data analyses
are proved to be  robust, one would have to admit the possibility of a very large $\nue$ component in the flux before neutrino conversion (i.e. $\phi_e>\phi_x$ above the detection threshold), or  invoke exotic physics
that could give a larger $\barnue$ survival probability, $\bar
p_{max}$ (see Eq. (\ref{magg_2})). 
A measurement above the limit (\ref{unitlim}) would exclude the adiabatic inverted scenario with $\nux$ dominance.
A $\nue$ detection would also constrain the $\barnue$ flux indirectly.
For example, Eq. (\ref{bestnuenum}) (with ${\rm Max}(\phi_{\bar x}/\phi_{ x})=1.5$) tells that a measured $\nue$ flux smaller than $\sim 0.48~ {\rm cm^{-2} s^{-1}} $ above 19.3 MeV would be incompatible with a $\barnue$ flux as large as $1.2~ {\rm cm^{-2} s^{-1}}$ in the same energy bin, thus improving the SK bound for $\barnue$, Eq. (\ref{sklim}). 

In summary, I found an upper limit of $\Phi_e \lta 5.5 ~{\rm
  cm^{-2} s^{-1}}$ above 19.3 MeV   for the $\nue$ component of the
diffuse supernova neutrino flux,  using the SK bound on the $\barnue$ flux, the  properties of supernova neutrino spectra and neutrino oscillations.  This new limit is the  strongest  available,
improving the  LSD bound by three orders of
magnitude.  It defines the necessary sensitivity for future
searches of the $\nue$ flux.  Noticeably, the result proves that the $\nue$ flux needs not
be  equal or smaller than the $\barnue$ one -- as it would be the case without
oscillations -- but instead it can be larger by a factor of several
above $E\sim 20$ MeV, depending on the oscillation scenario.  It follows that, in spite of their smaller volumes,
$\nue$ telescopes can be competitive with SK in the search of the
diffuse supernova neutrino flux.  A dedicated
analysis of the existing SNO data is compelling, in this perspective.

I am very grateful to O.~L.~G.~Peres, A.~Yu.~Smirnov, and M.~Strassler  for
precious comments on the preliminary manuscript, and to M.~Maltoni for useful discussions.  I thank ICTP of Trieste for hospitality while working on this paper, and acknowledge
support from the INT-SCiDAC grant number DE-FC02-01ER41187.

\bibliography{draftPRLversion}

\begin{thebibliography}{99}


\bibitem{Malek:2002ns}
  M.~Malek {\it et al.},
  Phys.\ Rev.\ Lett.\  {\bf 90}, 061101 (2003).

\bibitem{Totani:1995dw}
  T.~Totani, K.~Sato and Y.~Yoshii,
  Astrophys.\ J.\  {\bf 460}, 303 (1996).

\bibitem{Malaney:1996ar}
  R.~A.~Malaney,
  Astropart.\ Phys.\  {\bf 7}, 125 (1997).

\bibitem{Hartmann:1997qe}
{D.~H.} {Hartmann}  {and}
  {S.~E.} {Woosley},
  {\it Astropart. Phys.} {{\bf 7}},
  {137} ({1997}).

\bibitem{Kaplinghat:1999xi}
{M.}~{Kaplinghat},
  {G.}~{Steigman}, {and}
  {T.~P.} {Walker},
  {\it Phys. Rev.} {{\bf D62}},
  {043001} ({2000}).

\bibitem{Ando:2004sb}
  S.~Ando,
  Astrophys.\ J.\  {\bf 607}, 20 (2004).

\bibitem{Ando:2004hc}
{S.}~{Ando}  {and}
  {K.}~{Sato}, {\it New
  J. Phys.} {{\bf 6}}, {170}
  ({2004}).

\bibitem{Strigari:2003ig}
{L.~E.} {Strigari}, {\it et al.},
  {\it JCAP} {{\bf 0403}}, {007}
  ({2004}).

\bibitem{Lunardini:2005jf}
  C.~Lunardini,
  astro-ph/0509233.

\bibitem{Hirata:1987hu}
  K.~Hirata {\it et al.},
  Phys.\ Rev.\ Lett.\  {\bf 58}, 1490 (1987).

\bibitem{Bionta:1987qt}
  R.~M.~Bionta {\it et al.},
  Phys.\ Rev.\ Lett.\  {\bf 58}, 1494 (1987).

 

\bibitem{Aglietta:1992yk}
  M.~Aglietta {\it et al.},
  Astropart.\ Phys.\  {\bf 1}, 1 (1992).

\bibitem{Beacom:2005it}
  J.~F.~Beacom and L.~E.~Strigari,
  hep-ph/0508202.

\bibitem{Cocco:2004ac}
  A.~G.~Cocco {\it et al.},
  JCAP {\bf 0412}, 002 (2004).
  
  \bibitem{Keil:2002in}
See e.g. {M.~T.} {Keil},
  {G.~G.} {Raffelt},
  {and} {H.~T.} {Janka},
  {\it Astrophys. J.} {{\bf 590}},
  {971} ({2003}).
  
\bibitem{Fukugita:2003en}
See e.g.  M.~Fukugita and T.~Yanagida,
  ``Physics of neutrinos and applications to astrophysics'',
Springer (2003) 593 p, and references therein.

\bibitem{Horowitz:2001xf}
  C.~J.~Horowitz,
  Phys.\ Rev.\ D {\bf 65}, 043001 (2002).
  
\bibitem{Totani:1997vj}
  T.~Totani {\it et al.}, 
  Astrophys.\ J.\  {\bf 496}, 216 (1998).

\bibitem{Burrows:2000tj}
  A.~Burrows {\it et al.},
  Astrophys.\ J.\  {\bf 539}, 865 (2000).

\bibitem{Liebendoerfer:2003es}
  M.~Liebendoerfer {\it et al.},
  Astrophys.\ J.\  {\bf 620}, 840 (2005).


\bibitem{Krastev:1988yu}
{P.~I.} {Krastev}  {and}
  {S.~T.} {Petcov},
  {\it Phys. Lett.} {{\bf B205}},
  {84} ({1988}).

\bibitem{Aharmim:2005gt}
  B.~Aharmim {\it et al.},
  Phys.\ Rev.\ C {\bf 72}, 055502 (2005).
  
\bibitem{Apollonio:1999ae}
  M.~Apollonio {\it et al.} 
  Phys.\ Lett.\ B {\bf 466}, 415 (1999).


  \bibitem{Dighe:1999bi}
{A.~S.}~{Dighe}  {and}
  {A.~Y.}~{Smirnov},
  {\it Phys. Rev.} {{\bf D62}},
  {033007} ({2000}).

\bibitem{Lunardini:2003eh}
{C.}~{Lunardini}  {and}
  {A.~Y.}~{Smirnov},
  {\it JCAP} {{\bf 0306}}, {009}
  ({2003}).

\bibitem{Lunardini:2001pb}
{C.~Lunardini}  {and}
  {A.~Y.~Smirnov},
  {\it Nucl. Phys.} {{\bf B616}},
  {307} ({2001}).


\bibitem{Mikheev:1986if}  S.~P.~Mikheev and A.~Y.~Smirnov,
  
Sov.\ Phys.\ JETP {\bf 64}, 4 (1986)  
[Zh.\ Eksp.\ Teor.\ Fiz.\  {\bf 91}, 7 (1986)]. 



\bibitem{Takahashi:2001dc}
  K.~Takahashi and K.~Sato,
  Phys.\ Rev.\ D {\bf 66}, 033006 (2002).


\bibitem{Schirato:2002tg}
  R.~C.~Schirato, G.~M.~Fuller,
  astro-ph/0205390.

\bibitem{Ando:2002zj}
  S.~Ando and K.~Sato,
  Phys.\ Lett.\ B {\bf 559}, 113 (2003).




  


\end{thebibliography}
\bibliographystyle{apsrev}

\end{document}